\RequirePackage{fix-cm}

\documentclass[smallcondensed]{svjour3}     

\smartqed  

\usepackage{graphicx}
\usepackage{subfig}
\usepackage{color}
\usepackage{verbatim}
\usepackage{amsmath}
\usepackage{amssymb}

\journalname{}
\begin{document}

\title{User Profiling Using Hinge-loss Markov Random Fields}
\subtitle{Inferring Age, Gender and Personality of Social Media Users\\
Using Textual, Visual and Relational Data}


\author{Golnoosh Farnadi\and Lise Getoor \and 
Marie-Francine Moens \and Martine De Cock
}


\institute{Golnoosh Farnadi \at
              Mila, Universit\'e de Montr\'eal, Canada\\
              \email{farnadig@mila.quebec} 
           \and
           Lise Getoor \at
           Statistical Relational Learning Group, University of California, Santa Cruz, USA\\
           \email{getoor@soe.ucsc.edu}
           \and
            Marie-Francine Moens \at
            Dept.~of Computer Science, Katholieke Universiteit Leuven, Belgium\\
            \email{sien.moens@cs.kuleuven.be} 
            \and
            Martine De Cock \at 
            Center for Data Science, University of Washington Tacoma, USA\\
            \email{mdecock@uw.edu} 
}

\date{}

\maketitle

\newcommand{\magenta}[1]{\textcolor{magenta}{#1}}
\newcommand{\martine}[1]{\magenta{\textsc{martine:} #1}}

\newcommand{\cyan}[1]{\textcolor{cyan}{#1}}
\newcommand{\golnoosh}[1]{\cyan{\textsc{golnoosh:} #1}}


\maketitle

\begin{abstract}

A variety of approaches have been proposed to automatically infer the profiles of users from their digital footprint in social media. Most of the proposed approaches focus on mining a single type of information, while ignoring other sources of available user-generated content (UGC). In this paper, we propose a mechanism to infer a variety of user characteristics, such as, age, gender and personality traits, which can then be compiled into a user profile. To this end, we model social media users by incorporating and reasoning over multiple sources of UGC as well as social relations. Our model is based on a statistical relational learning framework using Hinge-loss Markov Random Fields (HL-MRFs), a class of probabilistic graphical models that can be defined using a set of first-order logical rules. We validate our approach on data from Facebook with more than 5k users and almost 725k relations. We show how HL-MRFs can be used to develop a generic and extensible user profiling framework by leveraging textual, visual, and relational content in the form of status updates, profile pictures and Facebook page likes. Our experimental results demonstrate that our proposed model successfully incorporates multiple sources of 
information 
and outperforms competing methods that use only one source of information or an ensemble method across the different sources for modeling of users in social media.

\end{abstract}

%
%

\section{Introduction}
\label{sec:intro}

Web users actively generate content in various social media platforms. Modeling users by inferring their age and gender plays an important role in providing personalized services, viral marketing, recommender systems and tailored advertisements~\cite{nowson2006identity}. In addition to  age and gender, previous work in the field of psychology has highlighted the value of identifying the personality traits of users as an aid in building adaptive and personalized systems to provide rich and improved user experiences \cite{Oliveira:2011,tkalcic2015personality}. 

Various computational approaches for user profiling and inferring age, gender and personality traits based on user-generated content~(UGC) have been proposed in recent years~\cite{rangel2015overview,Rothe-ICCVW-2015,farnadi2016}; more details on related works are presented in Sections~\ref{sec:text}, \ref{sec:image}, and \ref{sec:relation}. Much of these efforts are aimed at finding novel techniques to infer user profiles using only one type of information, such as the user's textual posts. However, in many social media platforms, users can generate content in different modalities, such as textual content (e.g., status updates, blog posts, tweets, comments, etc.) and visual content (e.g., photo and video), while connecting with each other, i.e., creating relational content. A framework that leverages all available information about users can learn more accurate user profiles. This is especially useful for platforms where not every user generates the same type of information, and models trained based on one source of information fail to produce accurate user profiles. Examples include users who write status updates but never upload pictures, or users who join social media platforms only to consume knowledge and to relate with each other, rather than producing any textual or visual content themselves.

To address this, we propose a flexible user profiling framework that infers age, gender and Big Five personality traits using both UGC and social relational content. Our approach is based on a statistical relational framework with Hinge-loss Markov Random Fields (HL-MRFs)~\cite{bach:arxiv15}. 
In particular, we use Probabilistic Soft Logic~(PSL), a probabilistic programming language for defining HL-MRFs using weighted first-order logical rules, making them very expressive and suitable for modeling relational data like social network graphs. Recently, PSL has been successfully used for social network applications, with state-of-the-art results, such as sentiment analysis in social networks~\cite{west2014exploiting}, social trust propagation~\cite{huang2013flexible} and spam detection in social networks~\cite{fakhraei:kdd15}. 

Related work on combining multiple sources of user data have focused on fusion of features~\cite{cui2010multiple,you2016cross,sakaki2014twitter} known as \textit{early fusion} techniques, or use ensemble techniques to combine the results predicted from each source (i.e., majority voting or weighted majority voting)~\cite{sakaki2014twitter}, known as \textit{late fusion} techniques. Much of these efforts focus on combining various UGC sources and ignore social relational content to infer user profiles. Existing approaches which incorporate social relational content with UGC are mostly focused on finding informative features to combine them with the relational content features such as communities~\cite{huang2015multi,zhou2015multi}. 
The graphical structure of social media platforms is a rich source of information on user behaviour which, when leveraged properly, is very valuable for user profiling. There are a few related works in which social graphs have been used to combine various sources of data for collective classification purposes. The focus of these existing approaches is on tasks other than user profiling, such as the hybrid recommender system by Kouki et al.~\cite{kouki:recsys15} that combines multiple sources of user rating activities.
Our proposed user profiling framework not only incorporates social relational content to collectively predict user characteristics but also provides a mechanism to combine various other sources of user data for more accurate modeling of users.  

In addition to incorporating multiple sources of information, we infer multiple characteristics of the users at the same time. Using PSL makes our framework interpretable, and it is easy to add rules to the PSL model to infer a new characteristic or, similarly, remove one from the model. Furthermore, our approach is flexible and lends itself easily to incorporating other sources of information beyond the ones we consider in this paper. Finally, our technique works well with missing data. In particular, it does not require the availability of user data for all the information sources that are considered in the model. 

We evaluate our model on data from Facebook with more than 5,000 users. We use the users' textual posts, profile pictures and pages that they like to extract their profile by inferring their age, gender and personality traits. For personality traits, we use the traits of a widely accepted model, the Big Five personality model, consisting of the following five traits: Openness to experience, Conscientiousness, Extraversion, Agreeableness, and Neuroticism \cite{costa2008revised}. Our experimental results show that our proposed HL-MRFs model efficiently combines various sources of UGC to learn more accurate user profiles. To investigate whether the accuracy gain is due to the use of HL-MRFs or due to leveraging information from multiple sources simultaneously, we have trained a series of alternative models, including single-source logistic regression (LR) models, single-source HL-MRFs models, majority voting ensembles of the single-source models, and LR multi-source models. Our proposed HL-MRFs model outperforms them all. Our contributions include (1) a general and flexible framework to infer user characteristics from textual, visual, and relational social media content; (2) two interchangeable probabilistic graphical sub-models for user profiling based on user-item relations; (3) extensive experimental validation of the proposed models 
for predicting age, gender and personality traits of Facebook users based on their status updates, profile pictures and page likes.

The remainder of this paper is structured as follows: 
after reviewing the preliminaries of HL-MRFs and PSL in Section~\ref{sec:psl}, in Section~\ref{sec:textualvisual} we present our proposed model, including probabilistic graphical sub-models for inferring user characteristics from textual and visual content (Section \ref{SEC:GENMOD}) and 
sub-models
for user profiling based on user-item relations (Section~\ref{sec:relational}).
An extensive experimental validation and empirical comparison of our proposed model with alternative models using Facebook data is presented in Section~\ref{SEC:EVAL}. Finally, we provide promising future directions of this work 
and conclude in Section~\ref{sec:conclusion}.

\section{HL-MRF\lowercase{s} and PSL}
\label{sec:psl}

Hinge-loss Markov Random Fields~(HL-MRFs) are general classes of conditional probabilistic models with continuous values. HL-MRFs models are log-linear probabilistic models which enable efficient tractable inference. The distinguishing key of these models that make them tractable is the use of hinge-loss potentials of the form:  
\begin{equation}
 \phi_{j}(Y,X) = [\max(l_{j}(Y,X),0)]^{p}
\end{equation}

\noindent where $l_{j}(Y,X)$ is a linear function of sets of random variables $X$ and $Y$, and $p \in \{1,2\}$ allows to define linear or squared potentials. The variables in $X$ and $Y$ take on continuous values in the unit interval $[0,1]$. A HL-MRFs model defines a conditional probability density function over random variables $Y$ and conditioned on random variables $X$ using a set of $n$ hinge-loss potentials as follows:
\begin{equation}
\label{eq:pslprobability}
P(Y|X) = \frac{1}{Z(\lambda)}\exp \left(-\sum_{j=1}^n \lambda_{j} \phi_{j}(Y,X)\right)
\end{equation}

\noindent where $Z(\lambda) = \int_{Y} \exp [-\sum_{j=1}^n \lambda_{j} \phi_{j}(Y,X)]$ and the weights $\lambda_{j}$ capture the importance of each potential in the model.

Throughout this paper we use Probabilistic Soft Logic (PSL), a weighted first-order logical language to specify HL-MRFs models, that makes these powerful models in\-ter\-pre\-ta\-ble, flexible and expressive. A PSL model consists of a set of PSL rules of the form:

\begin{equation}
\lambda_{r}: \underbrace{B_1 \land B_2 \ldots \land B_m}_{r_{body}} \rightarrow \underbrace{H}_{r_{head}}
\end{equation}

\noindent where $B_1, B_2, \ldots, B_m$ and $H$ are \textit{predicates} or negated predicates. Each predicate is of the form $p(a_1, a_2, \penalty 0 \ldots, a_v)$ where $p$ is a \textit{predicate symbol}, and each argument $a_1, a_2, \ldots,a_v$ is either a constant or a variable. By instantiating all the variables with constants from their domains in rule $r$, we ground the rule. $\lambda_{r} \in \mathbb{R}^{+} \cup \{\infty\}$ is the weight of the rule $r$.


An interpretation $I$ is a mapping that couples a continuous value $I(x) \in [0,1]$ to each ground predicate $x$ which specifies its value.
The value of a ground rule in PSL is calculated based on {\L}ukasiewicz logic~\cite{klir1995fuzzy}. Conjunction $\wedge$ is interpreted by the {\L}ukasiewicz t-norm ($\tilde{\wedge}$), disjunction $\vee$ by the {\L}ukasiewicz t-conorm ($\tilde{\vee}$), and negation $\neg$ by the {\L}ukasiewicz negator ($\tilde{\neg}$), which are defined as follows: for $p, q \in [0,1]$ we have 
$p \,\tilde{\wedge}\, q = \max(0 , p+q -1), $
$p \,\tilde{\vee}\, q = \min(p+q , 1)$ and
$\tilde{\neg} p = 1 - p$. 
The $\, \tilde{} \,$ indicates the relaxation over Boolean values.
The distance to satisfaction of a ground PSL rule $r$ is defined as:
\begin{equation}
\label{eq:psldts}
l_r =  I(r_{body}) - I(r_{head})
\end{equation}

\begin{example}
\label{example:distance}
\noindent Consider the PSL rule as $
{\textit Is}(U,C) \land {\textit Friend}(U,V) \rightarrow   {\textit Is}(V,C)
$. Given interpretation $I$, we instantiate $U = Alice$, $C = yng$ and $V = Bob$. This instantiation results in the grounded PSL rule $r_g$ as follows:

\begin{equation*}
{\textit Is}(Alice,yng) \land {\textit Friend}(Alice,Bob) \rightarrow   {\textit Is}(Bob,yng)
\end{equation*}

Let $I({\textit Is}(Alice,yng))=1$, i.e.~Alice is young to degree 1, and $I({\textit Friend}(Alice,$ $Bob))$ $= 0.7$, i.e.~Alice and Bob are friends to degree 0.7, then 
to fully satisfy the ground PSL rule $r_g$, $I({\textit Is}(Bob,yng))$ should be at least $\max(0 , 1+0.7-1) = 0.7$. If $I({\textit Is}(Bob,young)) = 0.5$, then $l_{r_{g}} = 0.2$.
\end{example}

The goal of Maximum a Posteriori~(MAP) inference in a HL-MRFs model (PSL model) is to find the most likely values for the variables in $Y$, given values for $X$:

\begin{equation}
I_{MAP}(Y) =\textit{arg }max_{I(Y)} P(I(Y)|I(X))
\end{equation}

\begin{example}
The grounded PSL rule $r_g$ in Example~\ref{example:distance}, results a hinge-loss potential function in the HL-MRF as:

\begin{equation*}
\max({\textit Is}(Alice,yng) + {\textit Friend}(Alice,Bob) - 1 - {\textit Is}(Bob,yng)) 
\end{equation*}
\end{example}

Each of the $n$ rules in a PSL model induces a hinge-loss potential of the form~\ref{eq:pslprobability}, in which the loss function $l_j$ is defined through the distance to satisfaction of the rule as in~\ref{eq:psldts}. By Equation~\ref{eq:pslprobability} it follows that the goal of optimization is to minimize the weighted sum of the distances to satisfaction of all rules. 

\noindent Since Equation~(\ref{eq:pslprobability}) is log-concave in $Y$, MAP inference in HL-MRFs models is a convex optimization problem and can be solved exactly via convex optimization. In this paper, we follow Bach et al.'s proposal of using an alternating direction method of multipliers (ADMM) based method for MAP inference in HL-MRFs \cite{bach:arxiv15}. Using ADMM allow to perform this optimization
efficiently and in parallel which makes the inference scalable, fast and efficient.

%
%

\section{User Profiling Model}
\label{sec:textualvisual}
In this section, we present our HL-MRFs model for user profiling in social media. In Section \ref{SEC:GENMOD} we first present a generic model, as well as two instantiations of it, namely \textsc{PSL-TXT} for inferences based on text, and \textsc{PSL-IMG} for inferences based on images. As PSL is ideally suited for modeling relational data, we dedicate a separate section, Section \ref{sec:relational}, to two models for inferences based on user-item relations, namely \textsc{PSL-DIRECT} and \textsc{PSL-LATENT}. All these individual models are seamlessly combined together into a user profiling model, called \textsc{PSL-PROFILE}, in Section \ref{sec:fusionmodel}.


\subsection{\textbf{Generic model}}\label{SEC:GENMOD}
Our PSL models for inferring the value of \textbf{characteristic} $C$ of \textbf{user} $U$ using \textbf{source} $S$ rely on the following two rules:

\begin{eqnarray}
{\textit Predicts}(U,C,S) & \rightarrow &   {\textit Is}(U,C) \label{metarule1}\\
{\textit Is}(U,C) & \rightarrow & {\textit Predicts}(U,C,S) \label{metarule2}
\end{eqnarray}

\noindent
Rules (\ref{textrule1})-(\ref{textrule2}) are ground versions of these rules, in which ${\textit Is}(carol,ext)$ denotes that $carol$ is an extrovert, and ${\textit Predicts}$ $(carol,$ $ext,$ $txt)$ means that it is predicted from $carol$'s text that she is an extrovert. In general, rule (\ref{metarule1}) expresses that the predicted value of a user's characteristic $c$ based on content from a source $s$ is indicative of the user's true value for this characteristic. Vice versa,  rule (\ref{metarule2}) says that the characteristics of a user should show through in the content they create. 

For some users, the value of characteristic $c$ might be known, while for others it has to be inferred. Similarly, for some users, some content of source $s$ might be available, allowing the use of any of the known single-source techniques from Section \ref{sec:text} and Section \ref{sec:image} to predict a value of characteristic $c$ for the user. A PSL model consisting of rules (\ref{metarule1})-(\ref{metarule2}) can consume all this existing evidence and infer values for the missing user characteristics. More formally, as explained in Section \ref{sec:psl}, let $X$ be all the evidence, i.e., $X$ is the set of ground predicates such that $\forall x \in X$, $I(x)$ is known, and let $Y$ be the set of ground predicates such that $\forall y \in Y$, $I(y)$ is unknown. Let $O$ be the set of all users, consisting of the set $O^+_{c}$ of users with known characteristic $c$, and the set $O^-_{c}$ of all users with unknown characteristic $c$ (i.e., $O = O^+_{c} \cup O^-_{c}$). Therefore $\forall u \in O^+_{c}$, $Is(u,c) \in X$ and $\forall u \in O^-_{c}$, $Is(u,c) \in Y$. 
%
%
In addition, if there is content of source $s$ available for user $u$, then we can use existing single-source techniques to infer a value for ${\textit Predicts}(u,c,s)$, i.e., ${\textit Predicts}(u,c,s) \in X$. 
A PSL model consisting only of the rules (\ref{metarule1})-(\ref{metarule2}) for a single source produces results that are close to the external single-source approach for inferring user characteristics. The true potential is reached when combining ground versions of rules (\ref{metarule1})-(\ref{metarule2}) for multiple sources.

The domain of variable $C$, $D(C)$ consists of \textit{fem}, \textit{yng}, \textit{opn}, \textit{con}, \textit{ext}, \textit{agr}, and \textit{neu}. We consider text and images as sources, i.e., $S = \{txt, img\}$.  These domains can be straightforwardly extended to include other information sources and user characteristics as well. 



Our PSL model \textsc{PSL-TXT} consists of the 14 rules that are obtained when grounding the variable $S$ in (\ref{metarule1})-(\ref{metarule2}) with $txt$, and grounding the variable $C$ with any of its values from it's domain $D(C)$
. Grounded versions of the predicate ${\textit Predicts}(U,C,txt)$ are to assign values to users whose textual content is available. 
To compute these grounded predicates, we first extract textual features from users textual content, such as unigram features, dictionary based features, topics, writing style, etc. Next we classify them with a trained model (e.g., logistic regression). The model is trained over users in the dataset for which both textual content and labels for the user characteristic at hand are available. More details, including a motivation for the choice of logistic regression for the external single-source models, are given in Section \ref{SEC:EVAL}. Examples of ground rules of the model \textsc{PSL-TXT} are:
%
%
\begin{eqnarray}
{\textit Predicts}(Carol,ext,txt) & \rightarrow &   {\textit Is}(Carol,ext) \label{textrule1}\\
{\textit Is}(Carol,ext) & \rightarrow & {\textit Predicts}(Carol,ext,txt) \label{textrule2}
\end{eqnarray}



Similarly, our PSL model \textsc{PSL-IMG} consists of the 14 rules that are obtained when replacing the variable $S$ in (\ref{metarule1})-(\ref{metarule2}) by $img$, and replacing the variable $C$ by any of its values from it's domain $D(C)$
. To assign values to grounded predicates ${\textit Predicts}(U,C,img)$ for users whose profile picture is available, we first extract visual features, e.g., pixels, from the images and score them with an external single-source model trained on instances where both the profile picture and the value of the user characteristic at hand are available. We provide more details about the external single-source approach in Section \ref{SEC:EVAL}. Examples of ground rules from \textsc{PSL-IMG} are:
\begin{eqnarray}
{\textit Predicts}(Carol,fem,img) & \rightarrow &   {\textit Is}(Carol,fem) \label{imagerule1}\\
{\textit Is}(Carol,fem) & \rightarrow & {\textit Predicts}(Carol,fem,img) \label{imagerule2}
\end{eqnarray}




%
%

\subsection{\textbf{Relational model}}
\label{sec:relational}

Next we introduce two PSL models that make use of user-item relational information. The first, which we refer to as the direct collaborative model (PSL-DIRECT) is based on the idea that if two users like the same item, then if one of them has a specific characteristic, the other has the same characteristic. 
Let $Likes(U,P)$ model the user-item relation. Using this relation we define the PSL model consisting of the following rules:
\begin{eqnarray}
\label{eq:direct1}
{\textit Is}(U,C) \land {\textit Likes}(U,P) \land  {\textit Likes}(V,P)\rightarrow   {\textit Is}(V,C)\\
\label{eq:direct2}
\neg {\textit Is}(U,C) \land {\textit Likes}(U,P) \land  {\textit Likes}(V,P)\rightarrow   \neg {\textit Is}(V,C)
\end{eqnarray}

The second, which we refer to as the latent model (PSL-LATENT), is based on the characteristics of the items. To infer the characteristics of users, first we infer the hidden (latent) characteristics of the items. In this model we use latent variables to define the item characteristics, i.e., $Represents(P,C)$ indicates that item $P$ represents characteristic $C$. The value of $I(Represents(P,C))$ is unknown for all items and inferred with MAP inference. 
The usefulness of using hidden factors to define items has been studied in~\cite{mcauley2013hidden}. The corresponding PSL model consists of the rules (\ref{eq:latent1}), (\ref{eq:latent2}), (\ref{eq:latent3}), and (\ref{eq:latent4}): 
\begin{eqnarray}
\label{eq:latent1}
{\textit Is}(U,C) \land {\textit Likes}(U,P) \rightarrow   {\textit Represents}(P,C)\\
\label{eq:latent2}
\neg {\textit Is}(U,C) \land {\textit Likes}(U,P) \rightarrow  \neg {\textit Represents}(P,C)\\
\label{eq:latent3}
 {\textit Represents}(P,C) \land {\textit Likes}(U,P) \rightarrow  {\textit Is}(U,C)\\
\label{eq:latent4}
\neg {\textit Represents}(P,C) \land {\textit Likes}(U,P) \rightarrow  \neg {\textit Is}(U,C)
\end{eqnarray}

We may have an item $p$ that has no page like relation from $u \in O^+_{C}$. We therefore add to our PSL-LATENT model the rules (\ref{eq:base1}) and (\ref{eq:base2}) that include prior knowledge from the evidence. To initiate values for the characteristics of all items, we assign the average score (${\textit Average}(C)$) of each characteristic $C$ using the characteristics of users in $O^+_{C}$:
\begin{eqnarray}
\label{eq:base1}
{\textit Average}(C)  \land {\textit Item}(P)  \rightarrow   {\textit Represents}(P,C)\\
\label{eq:base2}
{\textit Represents}(P,C) \land {\textit Item}(P) \rightarrow  {\textit Average}(C) 
\end{eqnarray}

For instance, a fully grounded rule of~\ref{eq:base1} for a page $123$ and a characteristic $fem$ is as follows: ${\textit Average}(fem)  \land {\textit Item}(123)  \rightarrow   {\textit Represents}(123,fem)$. Let $I({\textit Average}(fem)) = 0.61$ and $I({\textit Item}(123)) = 1$, then 
$I({\textit Represents}(123,fem)) \geq 0.61$. To force the value of $I({\textit Represents}(123,fem))$ to be equal to $0.61$, grounded Rule~\ref{eq:base2} is needed.

%
%

\subsection{\textbf{Prior model}}

To include prior knowledge from the evidence in our model for all users, we define a set of preliminary rules to assign the average score (${\textit Average}(C)$) of each characteristic $C$ using the characteristics of users in $O^+_{C}$. This prior knowledge is modeled with the PSL rules~(\ref{eq:base3})-(\ref{eq:base4}). The output of using a PSL model with these rules is equivalent to an average baseline approach. We call this model PSL-PRIOR.

\begin{eqnarray}
\label{eq:base3}
{\textit Average}(C)  \land {\textit User}(U)  \rightarrow   {\textit Is}(U,C)\\
\label{eq:base4}
 {\textit Is}(U,C) \land {\textit User}(U) \rightarrow  {\textit Average}(C) 
\end{eqnarray}

In our relational models (i.e., PSL-DIRECT and PSL-LATENT), we may have a user $u$ who has no page likes. We therefore add the meta rules (\ref{eq:base3}) and (\ref{eq:base4}) to these models to initiate values for all users.

\subsection{\textbf{Combined model}}\label{sec:fusionmodel}

\begin{figure}[t!]
  \centering
      \includegraphics[width=1\textwidth]{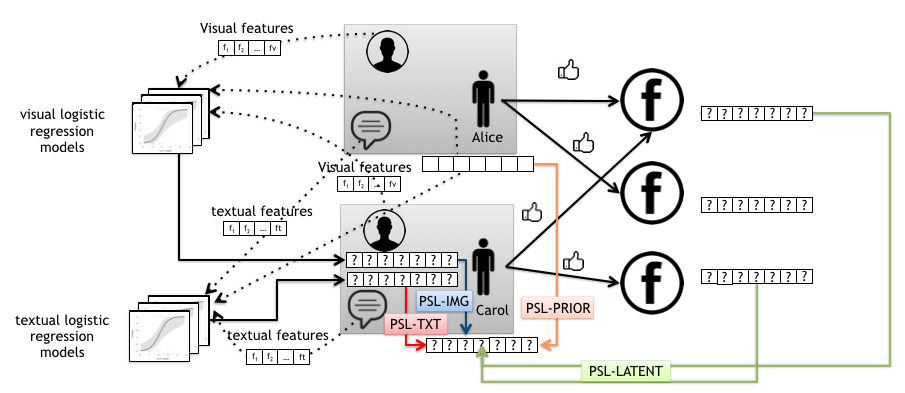}
  \caption{Architecture of the fusion model (PSL-PROFILE) by gathering information from text, image and page likes. In this model, Alice has known characteristics, while the characteristics of Carol are unknown. \label{fig:fusionmodel}}
\end{figure}

All the individual models that we have introduced can be combined together to build our PSL-PROFILE model to infer the unknown characteristics of users based on the known characteristics of other users using various sources of user data. Figure~\ref{fig:fusionmodel} presents the architecture of our PSL-PROFILE model, in which we combine PSL-PRIOR, PSL-TXT, PSL-IMG and PSL-LATENT models to infer the unknown characteristics of Carol based on the known characteristics of Alice. We extract textual features from the status updates of both Alice and Carol and train seven logistic regression models to predict the characteristics of Carol based on text, namely one model for each of the seven characteristics. We import these predicted values for each characteristic $C$ into our PSL-PROFILE model with the rules of the PSL-TXT model. Similarly, we extract visual features from the profile pictures of both Alice and Carol and train seven logistic regression models to predict the unknown characteristics of Carol using the known characteristics of Alice. We import the predicted results for each $C$ with the rules of PSL-IMG model into our PSL-PROFILE model. The pages that Alice and Carol like are modeled with the PSL-LATENT model to infer the known characteristics of Carol by using the hidden characteristics of pages that Carol likes. 
More details, including a motivation for the choice of the PSL-LATENT model for modeling user-item relations are given in Section \ref{SEC:EVAL}. 

\section{Empirical Evaluation}
\label{SEC:EVAL}
In this section, we present an experimental evaluation of our HL-MRFs models for user profiling on Facebook data. In Section \ref{SEC:DATASET}, we give details about the dataset from Facebook that we use in this study. To build a user profile, we incorporate three sources of user activities in Facebook, namely textual, visual and relational content. In Section~\ref{sec:text} the details on inferring age, gender and Big Five personality traits from status updates (i.e., textual information) are presented. Similarly, Section~\ref{sec:image} is dedicated to the details of predicting user characteristics from profile pictures (i.e., visual information). Next, in Section~\ref{sec:relation}, we present the evaluation of our two user-item models from Section~\ref{sec:relational} by comparing them with competing methods that leverage user page like information (i.e., relational information). All these sections start with an overview of related works in user profiling using each of the respective sources, i.e.~textual, visual, and relational. Finally, we present the complete evaluation results of applying our proposed HL-MRFs models on combined sources of information in Section~\ref{sec:finalResults}. 

\subsection{\textbf{Dataset and Evaluation Measures}}
\label{SEC:DATASET}

To validate all methods we use a subset of the MyPersonality project dataset\footnote{http://mypersonality.org/}. MyPersonality was a popular Facebook application introduced in 2007 in which users took a standard Big Five Factor Model psychometric questionnaire~\cite{questionnaire} and gave consent to record their responses and Facebook profile. The dataset contains information about each user's demographics, friendship links, Facebook activities (e.g., number of group affiliations, page likes, education and work history), status updates, profile picture and Big Five Personality scores. However, not all of this information is available for all users. We selected users who mention English as their language, and who provide age, gender, personality, status updates, page likes and a profile picture. 
To ensure that our prediction from the image belongs to the profile owner, we first selected profile pictures with only one face using OpenCV\footnote{http://opencv.org/} and a Haarcascade classifier~\cite{lienhart2002extended} and then re-selected the pictures with one face using the Project Oxford Face detector API\footnote{https://www.microsoft.com/cognitive-services/en-us/face-api}. 

By removing the Facebook pages with less than 3 likes by users in our dataset, our final dataset includes 49,372 pages, and
724,948 page like relations for 5,670 users. 
Personality traits are commonly described using five dimensions (known as the Big Five), i.e., Extraversion~(ext), Agreeableness~(agr), Conscientiousness~(con), Neuroticism~(neu), and Openness~(opn). The range of the personality scores in our dataset is between $[1,5]$. We use the median value to create binary classes for each characteristic, where median value for age = 23, opn = 4, con = 3.5, ext = 3.5, agr = 3.65, and neu = 2.75. 
We evaluate our user profiling model for the tasks of predicting age, gender and personality traits of Facebook users using their textual (status updates), visual (profile picture) and relational data (page likes). 

We make two sub-samples from our dataset, the first one is used for model tuning and the second one for testing our HL-MRF model. Our model tuning sub-sample is used in Section~\ref{sec:text}, Section
~\ref{sec:image} and Section~\ref{sec:relation} to select the single source textual, visual and relational models respectively. To measure the performance of the relational model in Section~\ref{sec:relation}, we extract the first sub-sample dataset for model tuning with those users from our dataset who liked more than 100 pages. Thus, the first sub-sample that we use as our model tuning dataset consists of 1,725 users, 48,641 Facebook pages and 592,907 page likes.

To build the second sub-sample, we use the remaining users, e.g., 3945 users, in our dataset as our test set. We systematically perform 10-fold cross-validation to collect all the results. The results of using the second sub-sample are presented in Section~\ref{sec:finalResults} and Section~\ref{sec:missingresults}. To gather the results, we use our second sub-sample as test set where 1,725 users of the first sub-sample are never used for testing the HL-MRF models. We perform 10-fold cross validation where the users of our first sub-sample are added as the training samples to the training set at each fold.

Since all the characteristics that we aim to predict are binary (i.e., positive class vs. negative class), to evaluate the results, we use the following metrics:
\textit{Accuracy}: the portion of correct results among all the test instances, \textit {AUC}: the area under the receiver operating characteristic curve, \textit{PR+}: the area under the precision-recall curve for the positive class, and \textit{PR-}: the area under the precision-recall curve for the negative class. The inferred results are turned into binary labels by mapping scores $<0.5$ to $0$ and scores $\geq 0.5$ to $1$. The natural language processing and machine learning approaches in the following sections are implemented using the scikit-learn library in Python. We implemented our HL-MRFs models using the publicly available PSL Java library with Groovy interface. Source code is available\footnote{ http://psl.umiacs.umd.edu}.

\subsection{\textbf{Textual Model Selection}}
\label{sec:text}

There is a substantial body of existing work on automatically inferring a user's characteristics from the user's digital footprint in social media platforms. Existing single-source models usually leverage either only text, image, or relational information. Machine learning models have been trained to infer the age, gender, and personality traits of users based on the \textit{textual context} they produce, including blog posts and status updates \cite{farnadi2016,rangel2015overview,schwartz2013personality}. Author profiling has gained a lot of attention in the past few years. Workshops and competitions such as PAN\footnote{http://pan.webis.de/} which focus on various features and techniques to predict age and gender of authors in various languages, or shared tasks such as WCPR\footnote{https://sites.google.com/site/wcprst/home/wcpr14} for personality prediction are a few examples.

During pre-processing, we combine the status updates of each user in the dataset into one document per user. From these, 
we extract two sets of textual features: (1) Linguistic Inquiry and Word Count (LIWC) features~\cite{Pennebaker1999} and (2) n-gram features. LIWC features are known to perform well in personality prediction \cite{farnadi2016}, and n-gram features are very popular and well-known in author profiling \cite{nowson2006identity}.

For each user, we extract 88 features using the LIWC tool, consisting of features related to (a) standard counts (e.g., word count), (b) psychological processes  (e.g., the number of anger words such as \textit{hate, annoyed, \ldots} in the text), (c) relativity (e.g., the number of verbs in the future tense), (d) personal concerns (e.g., the number of words that refer to occupation such as \textit{job, majors, \ldots}), and (e) linguistic dimensions (e.g., the number of swear words). For a complete overview, we refer to~\cite{Tausczik2010}.

\noindent(2) \textbf{n-grams}: for each user, we extract n-gram features where $n={1,2,3}$ from their status updates. As a weighting mechanism we use TF-IDF, and to select the $k$ top features we use Chi-square hypothesis testing where $k=2000$.



For each user characteristic, we train four models using the extracted LIWC features, namely a support vector machine with linear kernel classifier, a decision tree classifier, a Naive Bayes classifier, and a logistic regression classifier. 
And the logistic regression models outperform the other models for all characteristics in predicting the correct label. 
We train similar models over the n-gram features, 
where logistic regression again outperforms the other models. We omit the results because of space constraints. 
We then compare the performance of the LIWC-based models and n-gram-based models. 
The models based on the extracted LIWC features outperform the n-gram models in general. The n-gram-based model works slightly better than the LIWC based model to predict age and Neuroticism. As the textual predictor in our PSL models, e.g., PSL-TXT, we use the LIWC model trained with logistic regression.
Detailed results are presented in Table~\ref{tab:allPSLresults1}, Table~\ref{tab:allPSLresults2} and Table~\ref{tab:allPSLresults3} and discussed next. These include results about the performance of the LIWC based models with logistic regression as stand-alone single-source predictors in PSL-TXT, as well as when integrated into our PSL models including PSL-PROFILE.

\subsection{\textbf{Visual Model Selection}}
\label{sec:image}

Recently important progress has been made on age and gender identification from \textit{visual content} using deep neural networks. Rothe et al.,~for instance, successfully used a convolutional neural network (CNN) framework to detect the age and gender of users from their face \cite{Rothe-ICCVW-2015}. There are competitions concentrating on this task as well, such as the LAP Challenge 2016 on predicting apparent age estimation and gender classification of images\footnote{http://gesture.chalearn.org/}. Although much progress has been made in predicting age and gender from visual content, there is a limited focus on inferring personality traits. In~\cite{biel2013youtube}, Biel and Gatica-Perez focus on predicting personality of Vloggers (YouTube bloggers) based on their visual and audio content. Identifying personality traits from a static image such as a profile picture is mostly uncharted territory. Recently, in \cite{liu2016analyzing}, facial features (i.e., Face++ features) were extracted from Twitter profile pictures to predict personality, however, as the authors concluded, it is more challenging to predict a user's personality from a static image because less behavioural cues can be extracted from an image compared to other social media behaviours. In this paper we use a similar approach, based on Oxford project features.

For each user we use his/her profile picture. We extract 64 facial features from each profile picture using Microsoft Cognitive Services' Face API, also known as Project Oxford Face API~\cite{cao2010face}. 
The extracted features are face rectangle features to capture the location of the face in the image, face landmark features which include 27-point face landmarks pointing to the important positions of face components, face characteristics including age, gender, facial hair, smile, head position and glasses type. We refer to them as the ``Oxford features'' in the remainder of the paper.

Similar to the textual model, using the extracted Oxford features, we train a support vector machine with linear kernel, a decision tree, a Naive Bayes and a 
a logistic regression model per each user characteristic. The logistic regression models have the best overall performance. 

In addition to enabling the extraction of features from images, Project Oxford directly provides predictions for age and gender as well, making it a good alternative candidate for the external single-source predictor in \textsc{PSL-IMG}. However, for approx.~10\% of the users in our dataset (i.e.~543 out of 5,670), Project Oxford's native classifier doesn't produce a meaningful prediction. For this reason, in the remainder of this paper, we use our own logistic regression model trained over the Oxford facial features as the external single-source predictor from images in our PSL models. 
The age and gender prediction is more accurate using the Oxford API, with AUC score of 0.934 compared with 0.834 for the Oxford features trained with logistic regression model on our validation ser. 
Similarly, the AUC score of the Oxford API for the age prediction is 0.583 while using the Oxford features trained with logistic regression model give us AUC score of 0.523. 

%
It is important to note that all predictions in this paper are based on images that users have uploaded as their profile pictures. In the pre-processing step we filter single face pictures to enhance the chance of estimating the characteristics of the profile owner, but still this provides no guarantee that these images actually depict the profile owner. Many users upload pictures of their friends, family members or their child as their own profile picture, and therefore our predictions are an estimate of the characteristics of the face in the image and not necessarily the owner of the profile. 


\subsection{\textbf{Relational Model Selection}}
\label{sec:relation}

Existing work on inferring user characteristics from \textit{relational content} focuses typically either on using homophily or heterophily relations among friends~\cite{farnadi2015scalable,mcpherson2001birds}, or indirect relations among users such as shared Facebook page likes~\cite{mypersonality}. 

As HL-MRFs is ideally suited for modeling relational data, we defined two novel models, i.e., \textsc{PSL-DIRECT} and \textsc{PSL-LATENT}, for inferences based on user-item relations (see Section~\ref{sec:relational}). One would intuitively expect that the accuracy of these models grows with the amount of available page like information. To verify this, we extract a sub-sample dataset with those users from our dataset in Section \ref{SEC:DATASET} who liked more than 100 pages. To measure the impact on the predictive accuracy of the HL-MRFs models by changing the number of Facebook pages that a user likes, we make five sub-samples by randomly selecting 20, 40, 60, 80 and 100 page likes per user in our original sample.

Our sub-sample consists of 1,725 users, 48,641 Facebook pages and 592,907 page likes. Since in our sub-sample, for each user we have more than 100 likes, we can test the performance of changing the number of page likes from 20, 40, 60, 80 to 100 for the same set of users. 

\begin{table}[h!]
\caption{Characteristics of the five sub-sample datasets of 1,725 Facebook users with various page likes.}
\centering
\small{\begin{tabular}{l|c c c c c}
Model&20&40&60&80&100\\
\hline
\#pages& 17,108& 25,208 &30,868 & 34,921& 38,070\\
\#likes& 36,206& 70,686 & 105,166 & 139,646 &174,113\\
\end{tabular}}
\label{tab:relationsampledatasets}
\end{table}

\begin{figure}
\centering
\begin{tabular}{cc}
\subfloat[]{\includegraphics[width=0.5\linewidth]{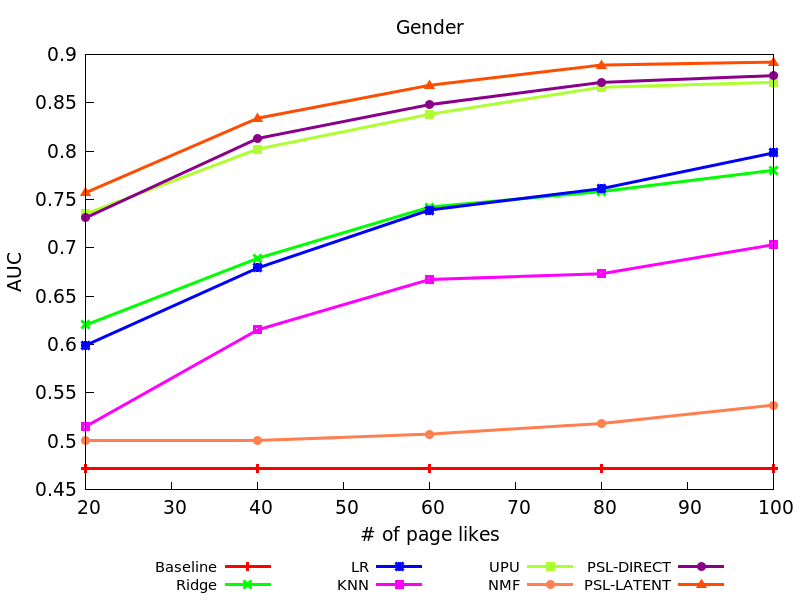}} &
\subfloat[]{\includegraphics[width=0.5\linewidth]{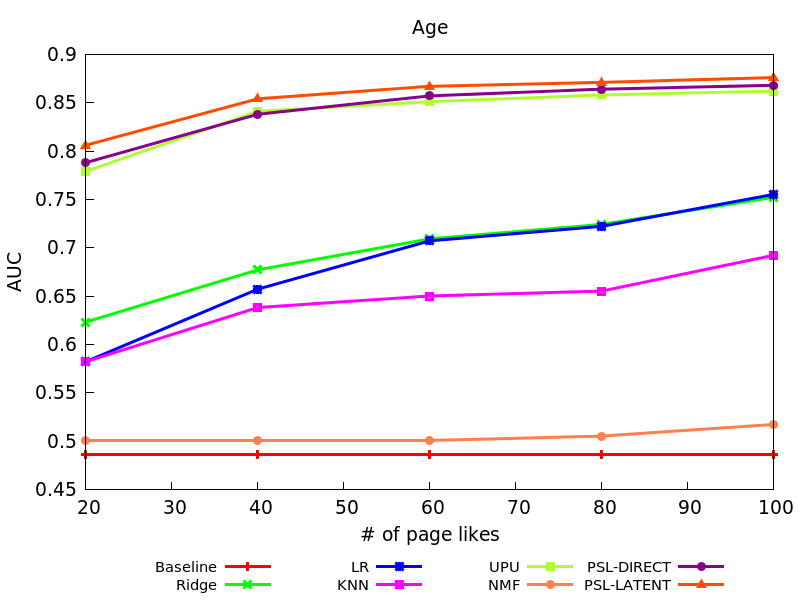}}\\
\subfloat[]{\includegraphics[width=0.5\linewidth]{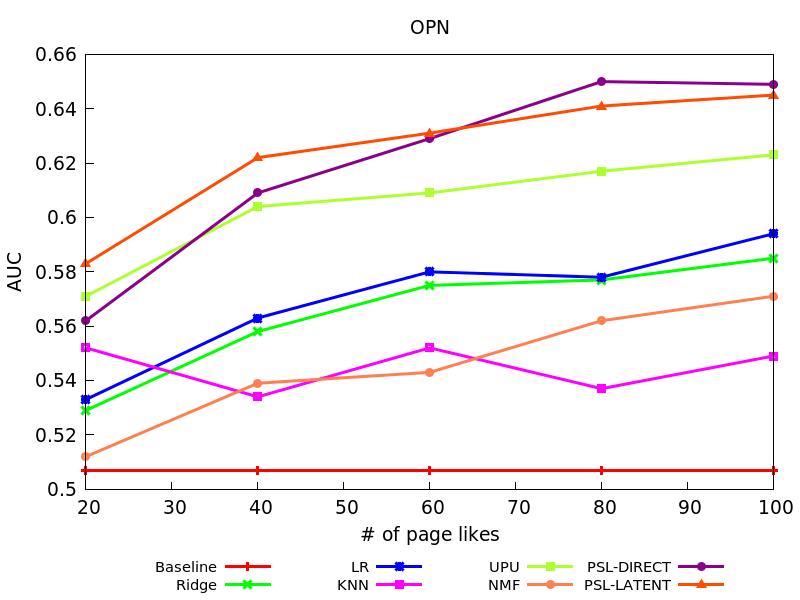}} &
\subfloat[]{\includegraphics[width=0.5\linewidth]{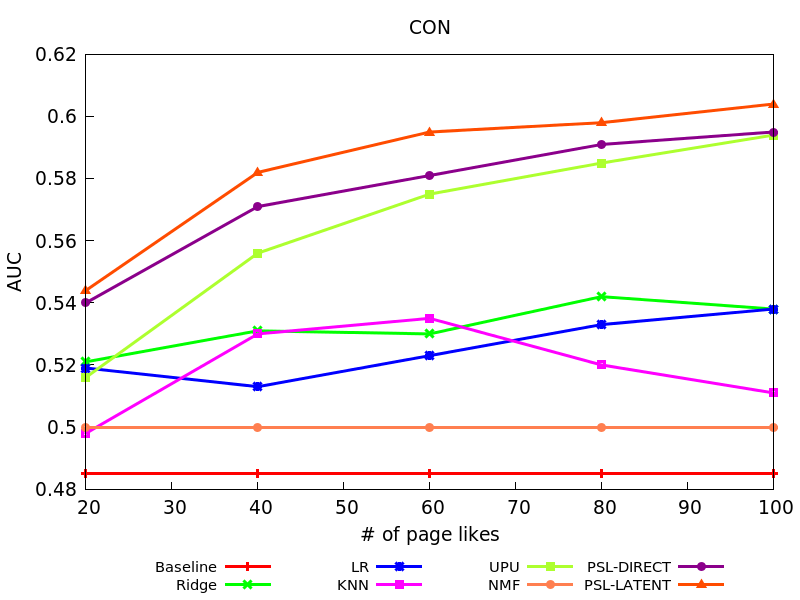}}\\
\subfloat[]{\includegraphics[width=0.5\linewidth]{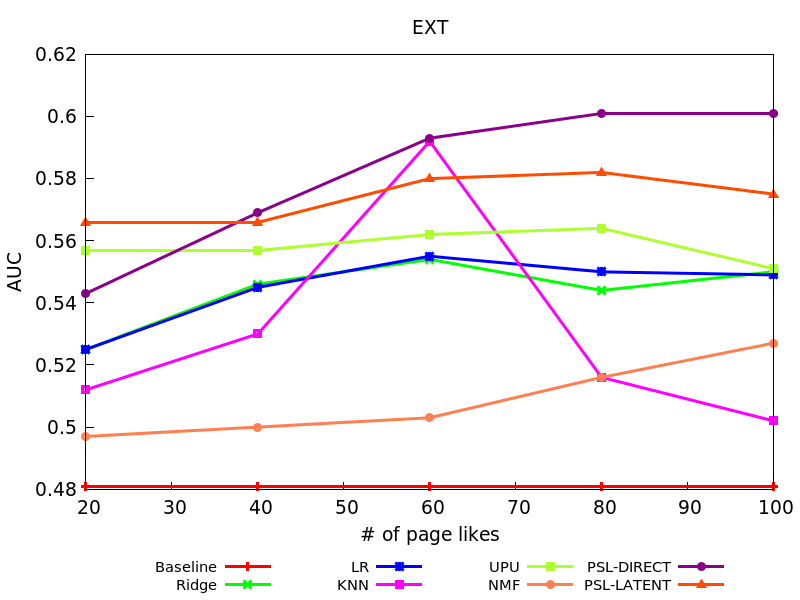}} &
\subfloat[]{\includegraphics[width=0.5\linewidth]{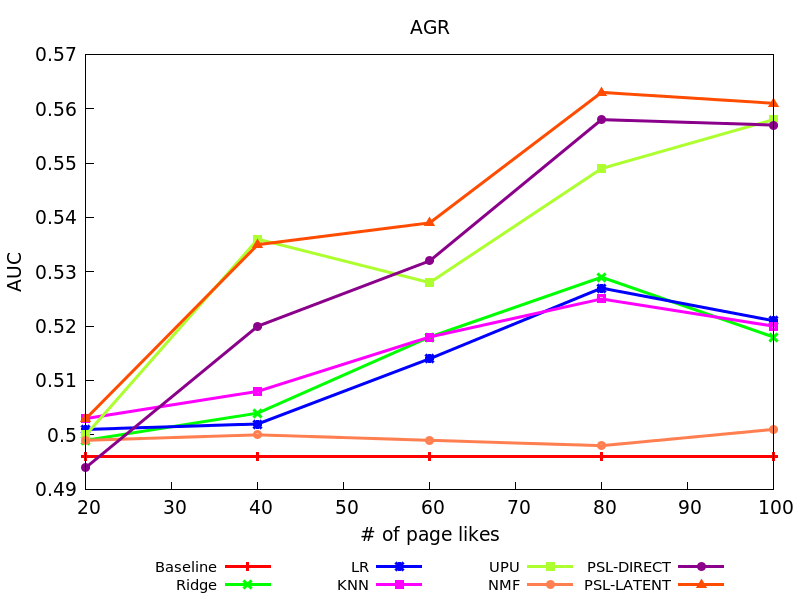}}\\
\multicolumn{2}{c}{\subfloat[]{\includegraphics[width=0.5\linewidth]{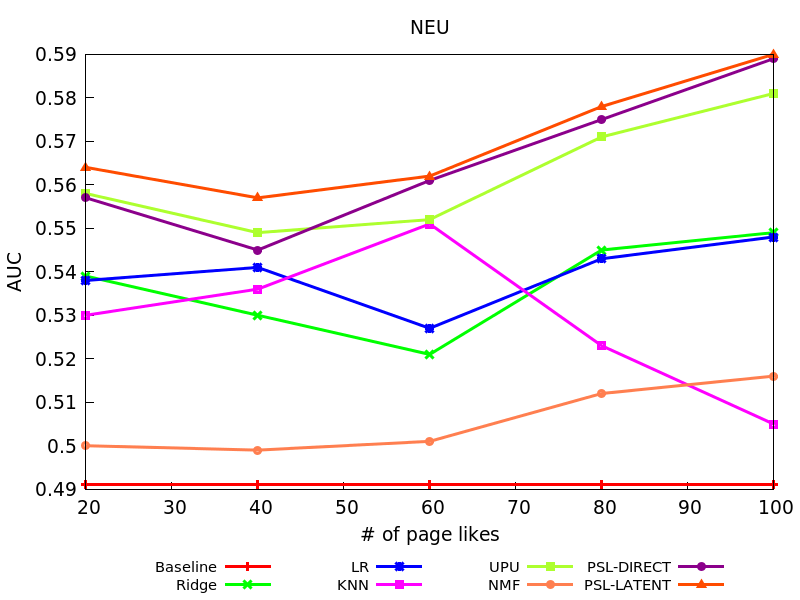}}} 
\end{tabular}
\caption{Area under the curve (AUC) results for predicting age, gender and personality traits based on three sub-sample datasets for user-item relations. 
\label{fig:relationalPSLresults}}
\end{figure}

We compare the performance of \textsc{PSL-DIRECT} and \textsc{PSL-LATENT} against the following approaches:

\noindent(1) \textbf{Average baseline}: We assign the average label from the training instances to the test instances. This baseline technique does not leverage page like information at all.
    
\noindent(2) \textbf{Matrix-based (Ridge)}: We use a matrix representation model as presented in~\cite{mypersonality}. In this model each row represents a user in the dataset and columns represent pages. The value of each matrix entry is one if the user likes that page in the dataset, otherwise it is zero. In \cite{mypersonality} Lasso is used to predict the Big Five personality traits, however since our labels are binary, we use a linear least squares classifier with $l2$ regularization (ridge regression). We set the parameter $\alpha$ to $0.1$.
    
\noindent(3) \textbf{Matrix-based (logistic regression (LR))}: Similar to the method mentioned above, we train a logistic regression classifier for each characteristic using the list of pages that each user likes. 
    
\noindent(4) \textbf{K-Nearest Neighbors (KNN)}: We find $k=5$ nearest neighbors of a given user based on the common pages that they like, and aggregate the labels with a majority vote.
    
\noindent(5) \textbf{User-Page-User (UPU)}: This approach is based on the graph structure and relies on a similar idea as our latent aggregate model (PSL-LATENT). In this approach, for each page we calculate the average score of the known characteristics of users who like that page. By aggregating the characteristics of users who like a page, we calculate the hidden characteristics of that page. Then, for a given user in our test set, we calculate the average score of the characteristics of the pages that a user likes using the pages' characteristics. 

\noindent(5) \textbf{Non-negative Matrix Factorization (NMF)}: We use non-negative matrix factorization to transform the user-item matrix and then predict the characteristics using logistic regression algorithm. Due to the sparsity of the matrix, neither changing the number of dimensions (from 2 to 7) nor the predictor enhance the performance of the NMF approach.

Figure~\ref{fig:relationalPSLresults} presents the results. Both of our proposed HL-MRFs models outperform other models in predicting users' characteristics. Both of our proposed models incorporate social relational data to collectively infer users' characteristics. Another interesting observation is that the PSL-LATENT model works much faster than the PSL-DIRECT model. The number of grounded rules (i.e., potentials) for both PSL-LATENT and PSL-DIRECT models depend on the number of page likes (i.e., items), however if we have $n$ page likes, the maximum number of potentials for the PSL-LATENT model per each characteristic is $4n$ while for the PSL-DIRECT model it is $2n^2$.


The PSL-LATENT model not only performs faster and more efficient than the direct model, but it also more accurately predicts the characteristics compared to the PSL-DIRECT model, except for Openness, Extroversion and Agreeableness where the differences between the PSL-LATENT model and the PSL-DIRECT model are not significant.

In all the models that we implemented for this study, increasing the number of page likes from 20 to 40, 60, 80 and then to 100 pages per user boosts the performance. Note that we are not using any information about the items, such as the name of the pages or their content. These results are in line with the results presented in~\cite{mypersonality}. Based on the results presented in Figure~\ref{fig:relationalPSLresults}, we select the latent model (i.e., PSL-LATENT) as the best relational model to capture the user-item relations in our combined user profiling models such as PSL-PROFILE discussed next.

\subsection{\textbf{User Profiling Results}}
\label{sec:finalResults}

To study the performance gains that can be achieved by using all sources of knowledge from text (status updates), image (profile pictures) and user-item relations (page likes),
%
we systematically make 10 folds by randomly splitting the users in our sample dataset. To have a fair comparison among the approaches that we have used, all the results in Table~\ref{tab:allPSLresults1}, Table~\ref{tab:allPSLresults2}, and Table~\ref{tab:allPSLresults3} are based on the same training and testing examples per each fold. The first line per each characteristic in all three tables (i.e., Table~\ref{tab:allPSLresults1}, Table~\ref{tab:allPSLresults2}, and Table~\ref{tab:allPSLresults3}), represents the average baseline (equivalent to PSL-PRIOR) results of that characteristic. 

\subsubsection{\textbf{Predictions based on one source}}

Using a single source of information, we present the results of the best model per each source in Table~\ref{tab:allPSLresults1}. The second line per each characteristic represents the results of the best \textit{textual model} using the users' status updates, i.e., where the extracted features are LIWC features and the trained model is logistic regression (PSL-TXT). The third line presents the results of the best \textit{visual model} using the users' profile picture by extracting the Oxford features and the trained model is logistic regression (PSL-IMG). And finally, the fourth line per each characteristic, shows the results of the best \textit{relational model} using HL-MRFs, which is the latent model (PSL-LATENT) that we have introduced in Section~\ref{sec:relational}. It is interesting that the relational models to predict users' age, gender and personality traits outperform the textual and visual models. Our textual models outperform our visual models for all characteristics except for gender where visual content performs significantly better than the textual model. All models using a single source of user data outperform the average baseline in predicting all characteristics. As expected, the performance of the PSL-IMG model in predicting personality traits from a single profile picture performs worse than the textual and relational content. 

\begin{table*}
\caption{Area under the curve (AUC), precision-recall curve for positive (PR+) and negative (PR-) results of using one source for inferring age, gender and personality traits. 
All results are averaged over a 10-fold cross-validation. Approaches are logistic regression (LR) and Hinge-loss Markov Random Fields (HL-MRFs). In each column, the highest determination are typeset in bold. 
}
\label{tab:allPSLresults1}
\centering
\renewcommand\arraystretch{1.1}
\renewcommand\tabcolsep{1pt}
\small{\begin{tabular*}{\linewidth}{@{\extracolsep{\fill}}l|c|c|c|ccc}
Characteristic&Source&Approach&Model&PR$^+$&PR$^-$&AUC\\
\hline
{Gender}&-&Baseline&PSL-PRIOR&0.593&0.407&0.497\\
&TXT&LR&PSL-TXT&0.689&0.553&0.645\\
&IMG&LR&PSL-IMG&0.828&\textbf{0.826}&0.845\\
&REL&HL-MRFs&PSL-LATENT&\textbf{0.885}&0.788&\textbf{0.856}\\
\hline
{Age}&-&Baseline&PSL-PRIOR&0.529&0.475&0.502\\
&TXT&LR&PSL-TXT&0.713&0.647&0.717\\
&IMG&LR&PSL-IMG&0.609&0.517&0.578\\
&REL&HL-RFs&PSL-LATENT&\textbf{0.875}&\textbf{0.840}&\textbf{0.868}\\
\hline
Openness&-&Baseline&PSL-PRIOR&0.420&0.556&0.488\\
&TXT&LR&PSL-TXT&0.502&0.603&0.565\\
&IMG&LR&PSL-IMG&0.448&0.561&0.502\\
&REL&HL-MRFs&PSL-LATENT&\textbf{0.582}&\textbf{0.685}&\textbf{0.652}\\
\hline
Conscientiousness&-&Baseline&PSL-PRIOR& 0.484&0.525&0.506\\
&TXT&LR&PSL-TXT&0.549&0.565&0.568\\
&IMG&LR&PSL-IMG&0.520&0.535&0.525\\
&REL&HL-MRFs&PSL-LATENT&\textbf{0.566}&\textbf{0.597}&\textbf{0.600}\\
\hline
Extroversion&-&Baseline&PSL-PRIOR&0.497&0.520&0.515\\
&TXT&LR&PSL-TXT&0.548&0.547&0.556\\
&IMG&LR&PSL-IMG&0.532&0.531&0.532\\
&REL&HL-MRFs&PSL-LATENT&\textbf{0.567}&\textbf{0.584}&\textbf{0.590}\\
\hline
Agreeableness&-&Baseline&PSL-PRIOR&0.504&0.505&0.507\\
&TXT&LR&PSL-TXT&0.553&0.534&0.554\\
&IMG&LR&PSL-IMG&0.538&0.519&0.533\\
&REL&HL-MRFs&PSL-LATENT&\textbf{0.567}&\textbf{0.540}&\textbf{0.567}\\
\hline
Neuroticism&-&Baseline&PSL-PRIOR&0.411&0.562&0.484\\
&TXT&LR&PSL-TXT&0.449&0.579&0.517\\
&IMG&LR&PSL-IMG&0.447&0.573&0.508\\
&REL&HL-MRFs&PSL-LATENT&\textbf{0.483}&\textbf{0.609}&\textbf{0.554}\\\\
\end{tabular*}}
\end{table*}

\subsubsection{\textbf{Predictions based on two sources}}
Next, we study the performance of the following combinations of two sources of information:

\noindent\textbf{(1) (Joint model) Textual+Visual}: To combine the textual and visual data, we extend the 88 LIWC features extracted from the status updates, with the 64 Oxford features extracted from the profile picture. We then train seven logistic regression models over these combined feature vectors. The second line in Table~\ref{tab:allPSLresults2} per each characteristic shows the results of this approach. This approach does not involve HL-MRFs models. This method of fusing textual and visual features has been widely used in the literature as an early multi-modal fusion technique. 

Another popular early technique for fusing features that has been proposed in the literature is canonical correlation analysis (CCA) \cite{correa2010canonical}. We investigated the use of CCA in user profiling by extracting the canonical correlation features for all pairs of textual, visual and relational contents (i.e., textual and visual, textual and relational, and visual and relational). Using the set of CCA features, we trained models for all the traits with logistic regression. We observed that the results of using CCA are worse than the joint model which simply combines the features. The main reason of getting poor results in using CCA for user profiling is that CCA does not utilize the users' traits in fusing the features from various sources; it only finds a linear combination of the existing variables for both knowledge sources such that the correlation is maximized for each source. Due to space limitations, the results of this CCA analysis are omitted from the paper. 


\noindent\textbf{(2) (HL-MRFs) Textual+Relational}: Since our relational model is not a matrix-based model, we use HL-MRFs to combine the relational model with the textual model. We implement this model in PSL using the rules~\ref{metarule1}, \ref{metarule2}, \ref{eq:latent1}, \ref{eq:latent2}, \ref{eq:latent3}, \ref{eq:latent4}, \ref{eq:base1} and \ref{eq:base2}. The results of applying this approach for each characteristic are shown in the third line (PSL-TXT+LATENT) in Table~\ref{tab:allPSLresults2}.

\noindent\textbf{(3) (HL-MRFs) Visual+ Relational}: Similar to the above combination, we use HL-MRFs to combine the visual and relational data. We implement this model in PSL using the same set of meta rules, however we change the source $S$ to $Img$ to use the predicted results from the profile pictures.  The last line for each characteristic in Table~\ref{tab:allPSLresults2} presents the results of this model (PSL-IMG+LATENT).

All models using two sources of data outperform the average baseline in predicting all characteristics. Beyond this, there are two key observations. The first important observation is that none of the approaches that combine textual and visual information perform particularly well. The joint LR approach does not outperform the single-source textual and visual predictors in predicting any of the traits. Our textual predictor outperforms our visual predictor for predicting age and personality traits, and also outperforms our joint technique. Even for the task of gender prediction where both textual and visual content perform reasonably well, the performance of the joint technique is not better than the visual predictor. The second important observation is that, in contrast to the above, combinations with relational data turn out to be successful. In particular, the HL-MRFs model that combines relational data with textual data not only outperforms the other techniques but also outperforms all single-source predictors in Table~\ref{tab:allPSLresults1}. The results of the HL-MRFs model that combines relational data with visual data are either better or as good as the single-source predictors in Table~\ref{tab:allPSLresults1}.  

\begin{table*}
\caption{Area under the curve (AUC), precision-recall curve for positive (PR+) and negative (PR-) results of using two sources for inferring age, gender and personality traits. 
All results are averaged over a 10-fold cross-validation. Approaches are logistic regression (LR) and Hinge-loss Markov Random Fields (HL-MRFs). In each column, the highest determination are typeset in bold. 
}
\label{tab:allPSLresults2}
\centering
\renewcommand\arraystretch{1.1}
\renewcommand\tabcolsep{1pt}
\small{\begin{tabular*}{\linewidth}{@{\extracolsep{\fill}}l|c|c|c|ccc}
Characteristic&Source&Approach&Model&PR$^+$&PR$^-$&AUC\\
\hline
{Gender}&-&Baseline&PSL-PRIOR&0.593&0.407&0.497\\
&TXT+IMG&LR&-&0.832&0.791 &0.842\\
&TXT+IMG&HL-MRFs&PSL-\{TXT+IMG\}&0.855&0.827&0.854\\
&REL+TXT&HL-MRFs&PSL-\{TXT+LATENT\}&0.871&0.773&0.849\\
&REL+IMG&HL-MRFs&PSL-\{IMG+LATENT\}& \bf{0.920}& \bf{0.886}& \bf{0.914}\\
\hline
{Age}&-&Baseline&PSL-PRIOR&0.529&0.475&0.502\\
&TXT+IMG&LR&-&0.709 &0.643 &0.712\\
&TXT+IMG&HL-MRFs&PSL-\{TXT+IMG\}&0.733&0.658&0.719\\
&REL+TXT&HL-MRFs&PSL-\{TXT+LATENT\}&\textbf{0.877}&\textbf{0.843}&\textbf{0.877}\\
&REL+IMG&HL-MRFs&PSL-\{IMG+LATENT\}&0.854&0.817&0.854\\
\hline
Openness&-&Baseline&PSL-PRIOR&0.420&0.556&0.488\\
&TXT+IMG&LR&-& 0.489 & 0.598 & 0.555\\
&TXT+IMG&HL-MRFs&PSL-\{TXT+IMG\}&0.494&0.603&0.563\\
&REL+TXT&HL-MRFs&PSL-\{TXT+LATENT\}&\textbf{0.575}&\textbf{0.681}&\textbf{0.650}\\
&REL+IMG&HL-MRFs&PSL-\{IMG+LATENT\}&0.558&0.660&0.630\\
\hline
Conscientiousness&-&Baseline&PSL-PRIOR& 0.484&0.525&0.506\\
&TXT+IMG&LR&-& 0.547 & 0.558 & 0.561\\
&TXT+IMG&HL-MRFs&PSL-\{TXT+IMG\}&0.553&0.553&0.560\\
&REL+TXT&HL-MRFs&PSL-\{TXT+LATENT\}&\textbf{0.583}&\textbf{0.598}&\textbf{0.607}\\
&REL+IMG&HL-MRFs&PSL-\{IMG+LATENT\}&0.565&0.579&0.588\\
\hline
Extroversion&-&Baseline&PSL-PRIOR&0.497&0.520&0.515\\
&TXT+IMG&LR&-& 0.546 & 0.545 & 0.555\\
&TXT+IMG&HL-MRFs&PSL-\{TXT+IMG\}&0.552&0.553&0.560\\
&REL+TXT&HL-MRFs&PSL-\{TXT+LATENT\}&\textbf{0.569}&\textbf{0.583}&\textbf{0.592}\\
&REL+IMG&HL-MRFs&PSL-\{IMG+LATENT\}&0.561&0.566&0.575\\
\hline
Agreeableness&-&Baseline&PSL-PRIOR&0.504&0.505&0.507\\
&TXT+IMG&LR&-& 0.550 & 0.558 & 0.561\\
&TXT+IMG&HL-MRFs&PSL-\{TXT+IMG\}&0.564&0.544&0.564\\
&REL+TXT&HL-MRFs&PSL-\{TXT+LATENT\}&\textbf{0.575}&\textbf{0.547}&\textbf{0.578}\\
&REL+IMG&HL-MRFs&PSL-\{IMG+LATENT\}&0.566&0.566&0.575\\
\hline
Neuroticism&-&Baseline&PSL-PRIOR&0.411&0.562&0.484\\
&TXT+IMG&LR&-& 0.456 & 0.584 & 0.525 \\
&TXT+IMG&HL-MRFs&PSL-\{TXT+IMG\}&0.465&0.588&0.527\\
&REL+TXT&HL-MRFs&PSL-\{TXT+LATENT\}&0.486&0.613&0.559\\
&REL+IMG&HL-MRFs&PSL-\{IMG+LATENT\}&\textbf{0.496}&\textbf{0.614}&\textbf{0.560}\\\\
\end{tabular*}}
\end{table*}

\subsubsection{\textbf{Predictions based on three sources}}

Finally, we investigate and compare the performance of the following models using all the three sources of users' data:

\noindent(1) \textbf{Ensemble (linear model)}: We predict the final characteristics of a user by getting the best prediction results from each source (i.e., the single source models) and apply majority voting. The second line per each characteristic in Table~\ref{tab:allPSLresults3} presents the results of this ensemble method, also known as a \textit{late fusion} approach. Most of the related works on combining various sources of UGC, which mostly focus on text and images and not on relations, are based on a linear combination of the predictions. Some papers propose to use a weighted average instead of averaging, such as~\cite{sakaki2014twitter}. There is no existing work on combining textual, visual and relational content. 
    
\noindent(2) \textbf{Joint model}: We extend our ``Textual+Visual" model with the relational features, i.e.~we extend the feature space of 88 LIWC features and 64 Oxford feature with the pages that each user likes. Since we have 49,372 pages in our sample, our page like matrix is very sparse. Therefore we first apply a truncated singular value decomposition (SVD) to reduce the dimensions and then extend it with our textual and visual features. We use logistic regression to train the models. The results of using this model are presented in the third line of each characteristic in Table~\ref{tab:allPSLresults3}. This technique is known as \textit{early fusion} approach.

    
\noindent(3) \textbf{PSL-PROFILE model}: We use our HL-MRFs fusion model to combine the predicted results of the visual and textual model with our latent model that leverages the page likes. The last line for each characteristic in Table~\ref{tab:allPSLresults3} contains the results of this approach. The architecture of this model is presented in Figure~\ref{fig:fusionmodel}.

\begin{table*}
\caption{Area under the curve (AUC), precision-recall curve for positive (PR+) and negative (PR-) results of using three sources for inferring age, gender and personality traits. 
All results are averaged over a 10-fold cross-validation. Approaches are logistic regression (LR) and Hinge-loss Markov Random Fields (HL-MRFs). In each column, the highest determination are typeset in bold. 
}
\label{tab:allPSLresults3}
\centering
\renewcommand\arraystretch{1.1}
\renewcommand\tabcolsep{1pt}
\small{\begin{tabular*}{\linewidth}{@{\extracolsep{\fill}}l|c|c|c|ccc}
Characteristic&Source&Approach&Model&PR$^+$&PR$^-$&AUC\\
\hline
{Gender}&-&Baseline&PSL-PRIOR&0.593&0.407&0.497\\
&REL+IMG+TXT&Ensemble&-&0.720&0.710&0.705\\
&REL+IMG+TXT&LR&-&0.838&0.798&0.845\\
&REL+IMG+TXT&HL-MRFs&PSL-PROFILE&\textbf{0.916}&\textbf{0.881}&\textbf{0.910}\\
\hline
{Age}&-&Baseline&PSL-PRIOR&0.529&0.475&0.502\\
&REL+IMG+TXT&Ensemble&-&0.663&0.643&0.660\\
&REL+IMG+TXT&LR&-&0.709&0.642&0.712\\
&REL+IMG+TXT&HL-MRFs&PSL-PROFILE&\textbf{0.880}&\textbf{0.827}&\textbf{0.876}\\
\hline
Openness&-&Baseline&PSL-PRIOR&0.420&0.556&0.488\\
&REL+IMG+TXT&Ensemble&-&0.527&0.589&0.567\\
&REL+IMG+TXT&LR&-&0.488&0.598&0.554\\
&REL+IMG+TXT&HL-MRFs&PSL-PROFILE&\textbf{0.588}&\textbf{0.674}&\textbf{0.649}\\
\hline
Conscientiousness&-&Baseline&PSL-PRIOR& 0.484&0.525&0.506\\
&REL+IMG+TXT&Ensemble&-&0.498&0.568&0.528\\
&REL+IMG+TXT&LR&-&0.551&0.560&0.564\\
&REL+IMG+TXT&HL-MRFs&PSL-PROFILE&\textbf{0.586}&\textbf{0.593}&\textbf{0.606}\\
\hline
Extroversion&-&Baseline&PSL-PRIOR&0.497&0.520&0.515\\
&REL+IMG+TXT&Ensemble&-&0.557&0.562&0.573\\
&REL+IMG+TXT&LR&-&0.546&0.545&0.555\\
&REL+IMG+TXT&HL-MRFs&PSL-PROFILE&\textbf{0.580}&\textbf{0.578}&\textbf{0.596}\\
\hline
Agreeableness&-&Baseline&PSL-PRIOR&0.504&0.505&0.507\\
&REL+IMG+TXT&Ensemble&-&0.542&0.545&0.552\\
&REL+IMG+TXT&LR&-&0.550&0.537&0.553\\
&REL+IMG+TXT&HL-MRFs&PSL-PROFILE&\textbf{0.578}&\textbf{0.552}&\textbf{0.582}\\
\hline
Neuroticism&-&Baseline&PSL-PRIOR&0.411&0.562&0.484\\
&REL+IMG+TXT&Ensemble&-&0.498&0.568&0.544\\
&REL+IMG+TXT&LR&-&0.456&0.585&0.526\\
&REL+IMG+TXT&HL-MRFs&PSL-PROFILE&\textbf{0.492}&\textbf{0.618}&\textbf{0.564}\\\\
\end{tabular*}}
\end{table*}

The results presented in Table~\ref{tab:allPSLresults3} indicate that our HL-MRFs based approach for combining the predictions from various sources not only allows the development of expressive and flexible models, but also outperforms the competing techniques. For all characteristics except for Extroversion and Conscientiousness, using our fusion model \textsc{PSL-PROFILE} based on all three sources of information significantly outperforms other combinations including results of using the single source and two sources of knowledge. For Extroversion and Conscientiousness, the outperforming model is based on the combination of the textual and relational models in HL-MRFs. For both characteristics, the predictive model based on the combination of all three sources works better for the positive class, and the combination of textual and relational models works better for the negative class. One important advantage of using our framework \textsc{PSL-PROFILE} is its ability to work with missing data, where we do not have all the information for all users. In this study, to have a fair comparison among the methods, we selected users who have all three knowledge sources, however our framework is directly applicable in a situation where not all users share similar content. In addition, our proposed framework can be used to create a more comprehensive user profile by gathering user data from different social media platform (e.g., Facebook and Twitter). It is important to note that combining only non-relational sources such as textual and visual content in HL-MRFs will produce results similar to an ensemble model. HL-MRFs models are ideally suited for modeling relational data and therefore they gain their power by combining non-relational sources with relational content.

%
%

\section{User Profiling Results With Missing Data}
\label{sec:missingresults}
A framework that leverages all available information about users can learn more accurate user profiles. This is especially useful for platforms where not every user generates the same type of information, and models trained based on one source of information fail to produce accurate user profiles. Examples include users who write status updates but never upload pictures, or users who join social media platforms only to consume knowledge and to relate with each other, rather than producing any textual or visual content themselves. In real-world social media platforms, users may not have a complete user profile. To mimic this behavior, in this section, we measure the performance of our framework \textsc{PSL-PROFILE} when a fraction of users do not have all sources of knowledge in their profile. To study the effects of null values, we design three set of experiments, in each we remove one source of data from a fraction of users and evaluate our \textsc{PSL-PROFILE} performance. We randomly removed $20\%$, $40\%$, $60\%$, and $80\%$ of textual, visual and relational content from the profile of users in our test set. For the case of $100\%$ missing values, the results are equivalent to the results presented in Table~\ref{tab:allPSLresults2} when we present the results of combining two sources of knowledge per each user in our dataset.

\begin{figure}
\centering
\begin{tabular}{cc}
\subfloat[]{\includegraphics[width=0.5\linewidth]{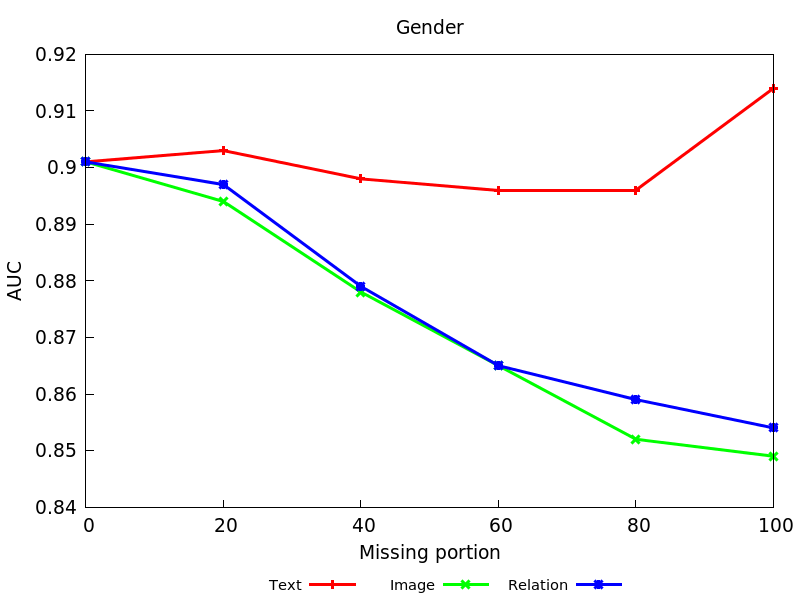}} &
\subfloat[]{\includegraphics[width=0.5\linewidth]{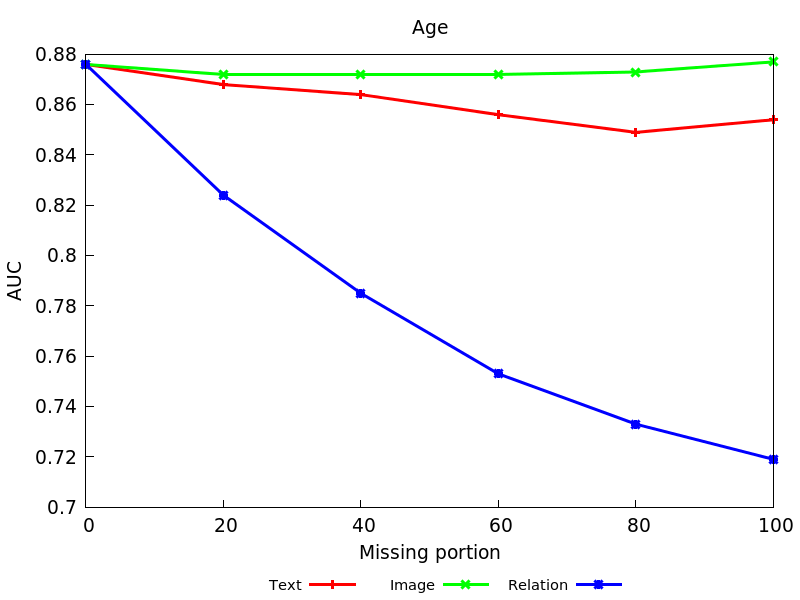}}\\
\subfloat[]{\includegraphics[width=0.5\linewidth]{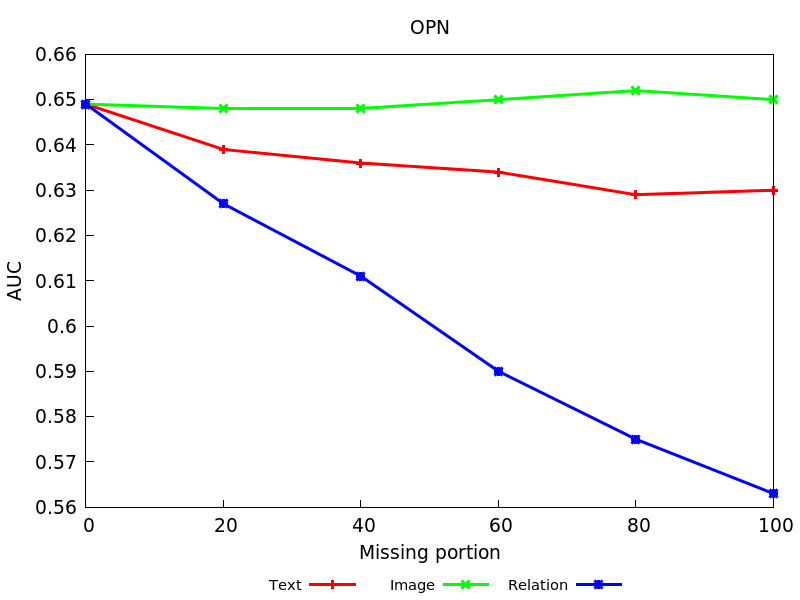}} &
\subfloat[]{\includegraphics[width=0.5\linewidth]{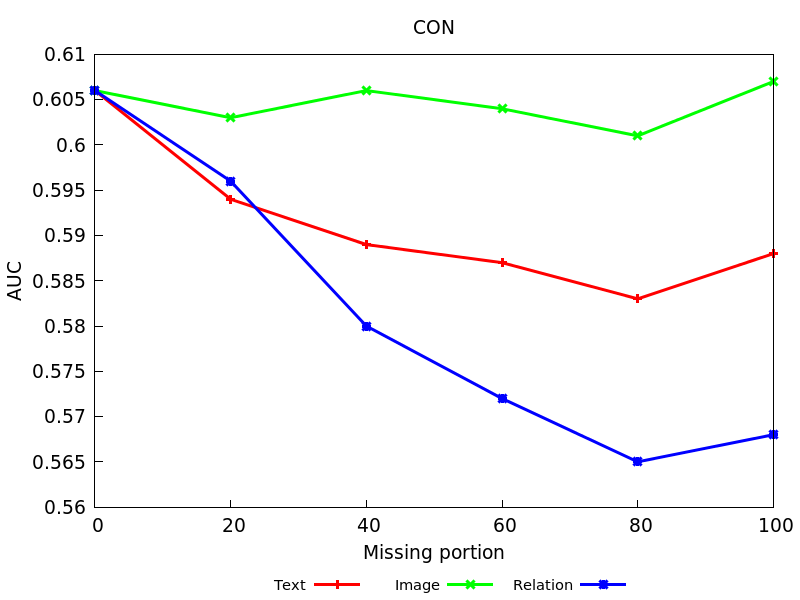}}\\
\subfloat[]{\includegraphics[width=0.5\linewidth]{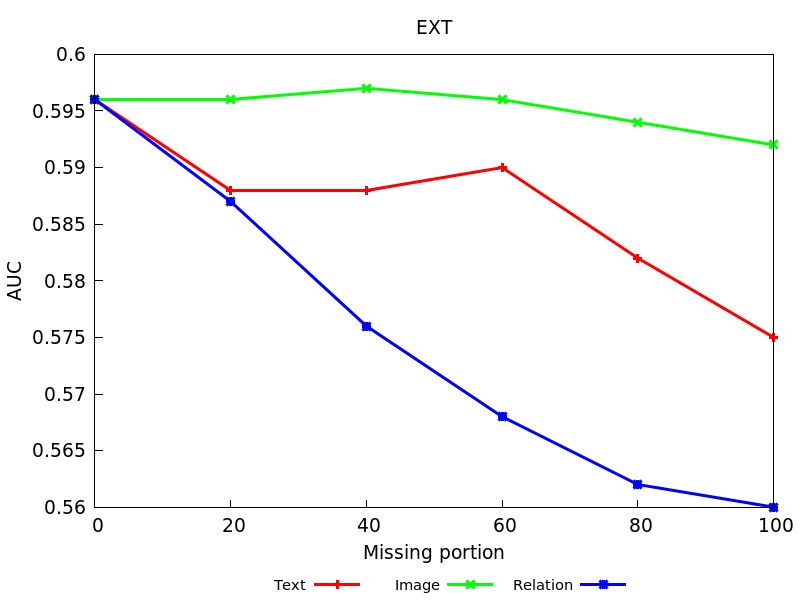}} &
\subfloat[]{\includegraphics[width=0.5\linewidth]{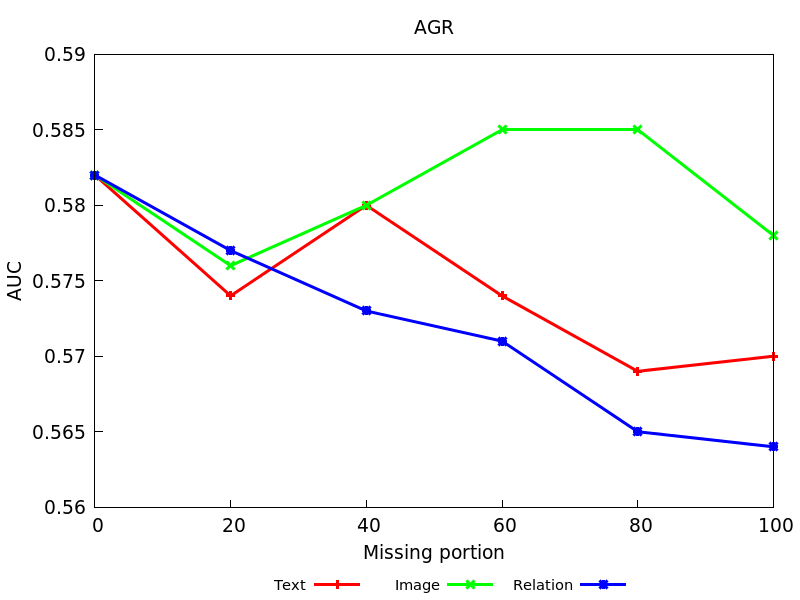}}\\
\multicolumn{2}{c}{\subfloat[]{\includegraphics[width=0.5\linewidth]{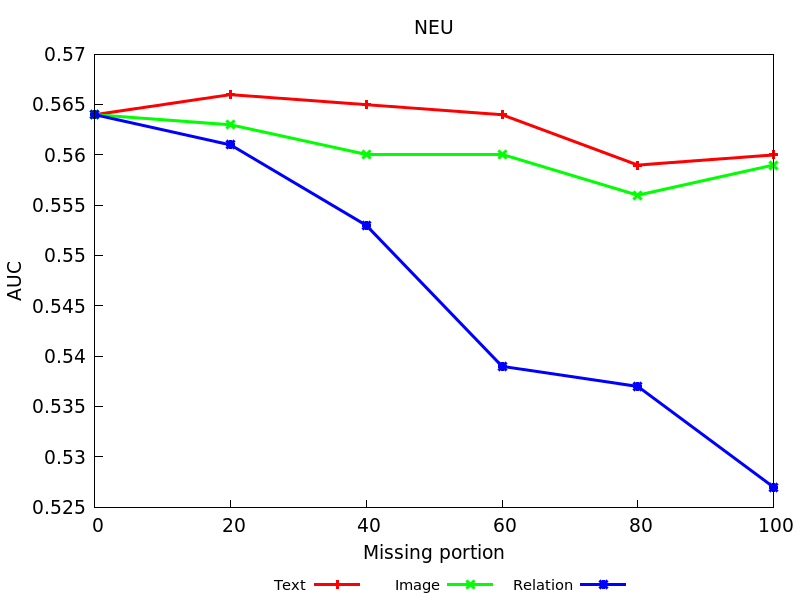}}} 
\end{tabular}
\caption{Area under the curve (AUC) results of age, gender and personality prediction with various percentage of missing data from text, image and relational data. All results are averaged over a 10-fold cross-validation. 
\label{fig:missingPSLresults}}
\end{figure}

The results presented in Figure~\ref{fig:missingPSLresults} indicate that except for the case of gender prediction that visual content is the best option, the most predictive data source for all the other characteristics is the relational content. Visual content for personality traits prediction and age prediction performs poorly, and missing the content has little influence on the overall results, however for the case of Neuroticism, textual content performs worst that the visual content. In all cases, missing data influence on the results and having more data enhance the prediction, unlikely missing the visual content for the case of age prediction has little influence on the overall results. Given the results presented in Figure~\ref{fig:missingPSLresults}, one can select data sources suitable for predicting each characteristics.

\section{Conclusion and Future Work}
\label{sec:conclusion}

In various social media platforms, users have the freedom to generate content in different modalities. Social media users can generate textual content in the form of a blog post, status updates, tweets, comments, etc. Similarly, they can upload or share photos and images as their profile picture. Although there are many research works that treat the various sources independently to infer user characteristics, there are not much related works that combine these predictive models effectively. 

In this paper, to model social media users, we combined various sources of user-generated and social relational content. We built a flexible and expandable user profiling model using a probabilistic graphical model, Hinge-loss Markov Random Fields, and used a probabilistic relational framework, Probabilistic Soft Logic (PSL) to implement it. We provided extensive experimental validation of the proposed model for predicting age, gender and personality of Facebook users based on their status updates, profile picture and page likes. Our experimental results show that compared to the results of competing methods that use only one source of information, jointly learning the characteristic, or an ensemble method across the different sources, the multimodal user profiling model that we proposed, provided significantly more accurate user profiles. 

As part of our future, we plan to extend our user profiling model to not only incorporate user-item relations but also integrate other social relational content such as direct user-user relations: friendship and follower links. We will evaluate our user profiling framework with homophily relations with PSL rules such as ${\textit Is}(U,C) \wedge   {\textit Friend}(U,V) \rightarrow  {\textit Is}(V,C)$ and 
$\neg {\textit Is}(U,C) \wedge   {\textit Friend}(U,V) \rightarrow \neg {\textit Is}(V,C)$. Another promising future direction is to incorporate a weight learning mechanism. By using weight learning not only can we determine the relative importance of each information source for predicting each trait, but also we can capture the quality of the content that each user produces in inferring their characteristics. For instance, we may have an accurate predictive model based on the textual content, however for Alice who has only one status update but many page likes, the prediction from her textual content should get a lower weight compared to the predictions from her page like relations. Exploring a way to tailor weights of each source with the quality of the content that each user produces is an open path to explore in the future.

\small{
\bibliographystyle{abbrv}
\bibliography{sigproc}}  

\begin{thebibliography}{10}

\bibitem{bach:arxiv15}
S.~H. Bach, M.~Broecheler, B.~Huang, and L.~Getoor.
\newblock {Hinge-Loss Markov Random Fields and Probabilistic Soft Logic}.
\newblock arXiv:1505.04406 [cs.LG], 2015.

\bibitem{biel2013youtube}
J.-I. Biel and D.~Gatica-Perez.
\newblock The youtube lens: Crowdsourced personality impressions and
  audiovisual analysis of vlogs.
\newblock {\em Proc. of IEEE Transactions on Multimedia}, 15(1):41--55, 2013.

\bibitem{cao2010face}
Z.~Cao, Q.~Yin, X.~Tang, and J.~Sun.
\newblock Face recognition with learning-based descriptor.
\newblock In {\em Proc. of IEEE Conference on Computer Vision and Pattern
  Recognition (CVPR)}, pages 2707--2714. IEEE, 2010.

\bibitem{correa2010canonical}
N.~M. Correa, T.~Adali, Y.-O. Li, and V.~D. Calhoun.
\newblock Canonical correlation analysis for data fusion and group inferences.
\newblock {\em IEEE signal processing magazine}, 27(4):39--50, 2010.

\bibitem{costa2008revised}
P.~T. Costa and R.~R. McCrae.
\newblock The revised {NEO} personality inventory {(NEO-PI-R)}.
\newblock {\em The SAGE Handbook of Personality Theory and Assessment},
  2:179--198, 2008.

\bibitem{cui2010multiple}
B.~Cui, A.~K. Tung, C.~Zhang, and Z.~Zhao.
\newblock Multiple feature fusion for social media applications.
\newblock In {\em Proc. of the ACM SIGMOD International Conference on
  Management of data}, pages 435--446. ACM, 2010.

\bibitem{fakhraei:kdd15}
S.~Fakhraei, J.~Foulds, M.~Shashanka, and L.~Getoor.
\newblock Collective spammer detection in evolving multi-relational social
  networks.
\newblock In {\em ACM SIGKDD Conference on Knowledge Discovery and Data
  Mining}, 2015.

\bibitem{farnadi2015scalable}
G.~Farnadi, Z.~Mahdavifar, I.~Keller, J.~Nelson, A.~Teredesai, M.-F. Moens, and
  M.~De~Cock.
\newblock {Scalable adaptive label propagation in Grappa}.
\newblock In {\em Proc. of IEEE International Conference on Big Data}, pages
  1485--1491. IEEE, 2015.

\bibitem{farnadi2016}
G.~Farnadi, G.~Sitaraman, S.~Sushmita, F.~Celli, M.~Kosinski, D.~Stillwell,
  S.~Davalos, M.-F. Moens, and M.~De~Cock.
\newblock Computational personality recognition in social media.
\newblock {\em User Modeling and User Adapted Interaction}, pages 1--34, 2016.

\bibitem{questionnaire}
L.-R. Goldberg, J.-A. Johnson, H.-W. Eber, R.~Hogan, M.-C. Ashton, C.-R.
  Cloninger, and H.-G. Gough.
\newblock The international personality item pool and the future of
  public-domain personality measures.
\newblock {\em Journal of Research in Personality}, 40(1):84--96, 2006.

\bibitem{huang2013flexible}
B.~Huang, A.~Kimmig, L.~Getoor, and J.~Golbeck.
\newblock A flexible framework for probabilistic models of social trust.
\newblock {\em {Social Computing, Behavioral-Cultural Modeling and
  Prediction}}, pages 265--273, 2013.

\bibitem{huang2015multi}
Y.~Huang, L.~Yu, X.~Wang, and B.~Cui.
\newblock A multi-source integration framework for user occupation inference in
  social media systems.
\newblock {\em World Wide Web}, 18(5):1247--1267, 2015.

\bibitem{klir1995fuzzy}
G.~Klir and B.~Yuan.
\newblock {\em {Fuzzy sets and fuzzy logic}}.
\newblock Prentice Hall New Jersey, 1995.

\bibitem{mypersonality}
M.~Kosinski, D.~Stillwell, and T.~Graepel.
\newblock Private traits and attributes are predictable from digital records of
  human behavior.
\newblock volume 110, pages 5802--5805. National Acad Sciences, 2013.

\bibitem{kouki:recsys15}
P.~Kouki, S.~Fakhraei, J.~Foulds, M.~Eirinaki, and L.~Getoor.
\newblock Hyper: A flexible and extensible probabilistic framework for hybrid
  recommender systems.
\newblock In {\em Proc. of ACM Conference on Recommender Systems}, 2015.

\bibitem{lienhart2002extended}
R.~Lienhart and J.~Maydt.
\newblock An extended set of haar-like features for rapid object detection.
\newblock In {\em Proc. of International Conference on Image Processing},
  volume~1, pages I--900. IEEE, 2002.

\bibitem{liu2016analyzing}
L.~Liu, D.~Preotiuc-Pietro, Z.~Riahi~Samani, M.~E. Moghaddam, and L.~Ungar.
\newblock Analyzing personality through social media profile picture choice.
\newblock In {\em Proc. of the International AAAI Conference on Web and Social
  Media}, 2016.

\bibitem{mcauley2013hidden}
J.~McAuley and J.~Leskovec.
\newblock Hidden factors and hidden topics: understanding rating dimensions
  with review text.
\newblock In {\em Proc. of the 7th ACM conference on Recommender Systems},
  pages 165--172. ACM, 2013.

\bibitem{mcpherson2001birds}
M.~McPherson, L.~Smith-Lovin, and J.~M. Cook.
\newblock Birds of a feather: Homophily in social networks.
\newblock {\em Annual Review of Sociology}, pages 415--444, 2001.

\bibitem{nowson2006identity}
S.~Nowson and J.~Oberlander.
\newblock The identity of bloggers: Openness and gender in personal weblogs.
\newblock In {\em Proc. of AAAI Spring Symposium: Computational Approaches to
  Analyzing Weblogs}, pages 163--167, 2006.

\bibitem{Pennebaker1999}
J.-W. Pennebaker and L.-A. King.
\newblock {L}inguistic styles: Language use as an individual difference.
\newblock {\em Journal of Personality and Social Psychology}, 77:1296--1312,
  1999.

\bibitem{rangel2015overview}
F.~Rangel, P.~Rosso, M.~Potthast, B.~Stein, and W.~Daelemans.
\newblock {Overview of the 3rd Author Profiling Task at PAN 2015}.
\newblock In {\em Proc. of CLEF}, 2015.

\bibitem{Oliveira:2011}
O.~Rodrigo~de, K.~Alexandros, C.~C. Pedro, V.~Ana Armenta Lopez~de, and N.~O.
\newblock Towards a psychographic user model from mobile phone usage.
\newblock In {\em Proc. of the International Conference on Human Factors in
  Computing Systems}, pages 2191--2196, 2011.

\bibitem{Rothe-ICCVW-2015}
R.~Rothe, R.~Timofte, and L.~Van~Gool.
\newblock Dex: Deep expectation of apparent age from a single image.
\newblock In {\em Proc. of ICCV, ChaLearn Looking at People workshop}, 2015.

\bibitem{sakaki2014twitter}
S.~Sakaki, Y.~Miura, X.~Ma, K.~Hattori, and T.~Ohkuma.
\newblock Twitter user gender inference using combined analysis of text and
  image processing.
\newblock {\em V\&L Net 2014}, page~54, 2014.

\bibitem{schwartz2013personality}
H.~A. Schwartz, J.~C. Eichstaedt, M.~L. Kern, L.~Dziurzynski, S.~M. Ramones,
  M.~Agrawal, A.~Shah, M.~Kosinski, D.~Stillwell, and M.~E. Seligman.
\newblock Personality, gender, and age in the language of social media: The
  open-vocabulary approach.
\newblock {\em PloS one}, 8(9):e73791, 2013.

\bibitem{Tausczik2010}
Y.-R. Tausczik and J.-W. Pennebaker.
\newblock The {P}sychological meaning of words: {LIWC} and computerized text
  analysis methods.
\newblock {\em Journal of Language and Social Psychology}, 29:24--54, 2010.

\bibitem{tkalcic2015personality}
M.~Tkalcic and L.~Chen.
\newblock Personality and recommender systems.
\newblock In {\em Recommender Systems Handbook}, pages 715--739. Springer,
  2015.

\bibitem{west2014exploiting}
R.~West, H.~S. Paskov, J.~Leskovec, and C.~Potts.
\newblock Exploiting social network structure for person-to-person sentiment
  analysis.
\newblock {\em Transactions of the Association for Computational Linguistics},
  2:297--310, 2014.

\bibitem{you2016cross}
Q.~You, J.~Luo, H.~Jin, and J.~Yang.
\newblock Cross-modality consistent regression for joint visual-textual
  sentiment analysis of social multimedia.
\newblock In {\em Proc. of the Ninth ACM International Conference on Web Search
  and Data Mining}, pages 13--22. ACM, 2016.

\bibitem{zhou2015multi}
X.~Zhou, W.~Wang, and Q.~Jin.
\newblock Multi-dimensional attributes and measures for dynamical user
  profiling in social networking environments.
\newblock {\em Multimedia Tools and Applications}, 74(14):5015--5028, 2015.

\end{thebibliography}

\end{document}